\documentclass[preprint,a4paper,floatfix,preprintnumbers,amsmath,amssymb]{revtex4}
\usepackage{graphicx}

\newcommand{\rv}{\boldsymbol{r}}

\begin{document}

\title{Instability and Vortex Rings Dynamics in a Three-Dimensional Superfluid Flow Through a Constriction}

\author{ F. Piazza$^{1,2}$, L. A. Collins$^3$, and Augusto Smerzi$^{1,2}$}

\affiliation
    {
      $^1$INO-CNR, BEC Center, Via Sommarive 14, 38123 Povo, Trento, Italy\\
      $^2$Dipartimento di Fisica, Universita' di Trento, 38123 Povo, Italy\\
      $^3$Theoretical Division, Mail Stop B214, Los Alamos National Laboratory, Los Alamos, New Mexico 87545
    }

\begin{abstract} 
We study the instability of a superfluid flow through a constriction in three spatial dimensions.
We consider a Bose-Einstein condensate at zero temperature in two different geometries: a straight waveguide and a torus. The constriction consists of a broad, repulsive penetrable barrier. In the hydrodynamic regime, we find that the flow becomes unstable as soon as the velocity at the classical (Thomas-Fermi) surface equals the sound speed inside the constriction. At this critical point, vortex rings enter inside the bulk region of the cloud. 
The nucleation and dynamics scenario is strongly affected by the presence of asymmetries in the velocity and density of the background condensate flow.
\end{abstract}

\maketitle

{\bf Introduction}. 

Even when flowing past an obstacle, a superfluid can be stationary, since its excitations appear only under certain conditions \cite{Leggett_1999}. For instance, a homogeneous Bose-Einstein condensate (BEC) of density $n$ running through a weak obstacle, whose presence only slightly changes the cloud chemical potential, is energetically unstable towards phononic excitations only above a critical velocity $v_c$ equal to the sound speed $c\propto \sqrt{n}$, according to the Landau criterion. 
On the other hand, the critical velocity and the nature of the excitations change in the presence of a strong obstacle that considerably perturbs the homogeneous flow. Helium flowing through narrow channels becomes unstable at velocities much lower than predicted by the Landau criterion, which this time would correspond to rotonic excitations, since, as first predicted by Feynman, vortices are nucleated inside the channel and carry energy away from the superfluid through the phase-slip mechanism \cite{Anderson_1966,Avenel_1985,Varoquaux_2006,Packard_1998}. Feynman's energetics argument is analogous to the Landau criterion, and is based on the choice of vortex rings, instead of rotons, as the excitations coming at the critical velocity. His conjecture has been refined by Varoquaux \cite{Varoquaux_2006}, who used half-rings as the excitations to be nucleated inside the channel. In general, the type of excitation and the critical velocity depend on the flow configuration. 


In contrast to nearly-incompressible fluids like helium, atomic BECs, due to an exquisite control over the confining potential, offer a wide choice of flow geometries and obstacles, and also permit the modification of the effective dimensionality. Imaging techniques also provide a means of addressing the system on the healing length scale, at which vortex dynamics takes place.
The superfluid critical velocity in an elongated BEC can be studied for instance by
sweeping a laser beam through the cloud \cite{Ketterle_1999}. At the critical velocity, both vortices in an elongated trap \cite{Inouye2001} and solitons for tight confinement transverse to the flow \cite{Engels_2007} have been experimentally observed.
Another way to study the stability of the superflow would be to perturb a persistent current, which has been experimentally demonstrated with a BEC in a torus \cite{Phillips_2007}. Apart from being an hallmark of superfluidity, BECs in toroidal geometries and the study of their critical velocities have very important technological applications for sensing devices, like SQUIDs already realized with superfluid helium and superconductors.
BECs are often very well described by the mean-field Gross-Pitaevskii equation (GPE), which provides, unlike for helium, a reliable theoretical model to study the instability mechanism and its dynamics.
The nucleation of a vortex-antivortex pair at the critical velocity has been studied 
numerically in two and three dimensions with an obstacle moving through a uniform condensate \cite{Rica_1992}, and 
in a two-dimensional waveguide with an orifice \cite{Srivastava_1996}. Flow dissipation caused by the formation of solitons in one dimension
has been investigated in \cite{Pavloff2002}. In a previous work \cite{Piazza_2009}, we studied the scenario of superfluid flow dissipation in an effective two-dimensional toroidal geometry.

\begin{figure}
  \includegraphics[scale=0.6]{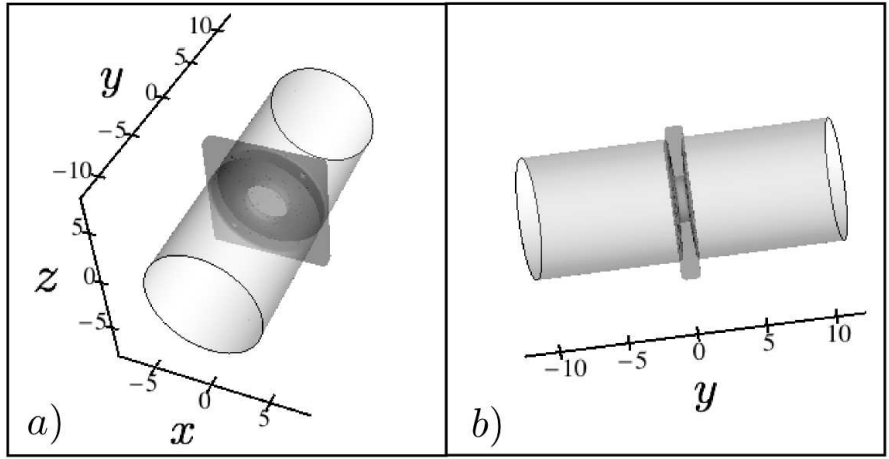}
  \caption {
        (color online). Constriction configuration. The light-gray surface corresponds to the classical (Thomas-Fermi) surface of the cloud. The dark-gray surface shows an isosurface of the barrier potential used to create the constriction.}
\label{wgsetup}
\end{figure}

In this manuscript, we study the critical velocity and superfluid flow dissipation mechanism in a three-dimensional constricted flow configuration, where the size of the cloud along all the three directions is much larger than the healing length $\xi=\hbar/\sqrt{2 m g n}$.
We consider a subsonic flow of a zero-temperature BEC in two different geometries: a wave guide with periodic boundary conditions, which can mimic an elongated cloud along the flow direction, and a torus. The unstable regime is reached by raising a repulsive penetrable barrier perpendicular to the flow.  This barrier is broader than the cloud dimension and extends over a few (typically 5 to 10) healing lengths along the flow direction. In this way, we create a constriction for the flow in the barrier region, as shown in Fig.~\ref{wgsetup}. Starting from a stationary flow with a given velocity, the barrier is adiabatically raised during the dynamical evolution until the instability sets in, and we observe vortex rings penetrating the cloud.
In this way, the superflow energy is dissipated through the phase-slip mechanism \cite{Anderson_1966}: vortices are able to take energy away from the flow since the velocity drops by a quantized amount in a region crossed by a vortex core \cite{Piazza_2009}.
The two geometries under study present significative differences in the vortex ring nucleation and dynamics. We find that vortex rings, which can find a stationary configuration after entering an axially symmetric waveguide, are instead always transient in the torus, and, more generally, as soon as the axial symmetry in the direction of flow is broken. 

{\bf Formulation and Computation}

In order to evolve these various cases, we solve a GPE (\ref{GP}) for three spatial dimensions in scaled form as
\begin{equation}\label{GP}
i \hbar \frac{\partial\psi(\boldsymbol{r},t)}{\partial t}=\left[-\frac{1}{2} \nabla^2 +V(\boldsymbol{r},t)+g|\psi|^2\right]\psi(\boldsymbol{r},t)
\end{equation}
where length, time, and energy are given in units of $d_o =  [\frac{\hbar}{m \omega_o}]^{\frac{1}{2}}$, $1/\omega_o$, and $\hbar \omega_o$ respectively for a representative
harmonic frequency $\omega_o$ that characterizes the trap.
 $V(\boldsymbol{r},t)$ is the external potential and $g=\frac{4\pi a}{d_o}$ with  $a$  the inter-particle $s$-wave scattering length and $m$  the atomic mass.  $\psi(\boldsymbol{r},t)$ is the condensate wavefunction normalized to the total number of particles $N$. The external potential has components associated with the
trapping and the barrier potentials of the form
\begin{equation}
V({\boldsymbol{r}},t) = V_{tr}({\boldsymbol{r}}) + V_b({\boldsymbol{r}},t) .
\end{equation}
In both cases,  we find the ground state of the system with $V_\mathrm{b}=0$.

The waveguide geometry is implemented by choosing the trapping potential
\begin{equation}
V_\mathrm{tr}(\rv)=\frac{1}{2} [x^2+ {\gamma}^2 z^2]\ ,
\label{wgtrap}
\end{equation}
with $\omega_o \equiv  \omega_x=30\times 2\pi \mathrm{Hz}$,  $\gamma = \omega_z/\omega_x$, and periodic boundary conditions along the flow direction $y$.  We considered three different values of the $\gamma$, namely 1, 1.05, and 1.2, which correspond, respectively, to a ground state chemical potential $\mu=11.7,\ 12.0,\ 16.5$ with $N=3\times 10^5$ $^{87}\mathrm{Rb}$ atoms and a nonlinear scaling value $g\ N=10134$ .

The toroidal geometry is implemented by choosing the trapping potential
\begin{equation}
V_\mathrm{tr}(\rv)=\frac{1}{2} [\alpha^2 x^2+ \beta^2 y^2 +  z^2] +V_c\ e^{-2(\rho/\sigma_c)^2}\ ,
\label{tortrap}
\end{equation}
where $\alpha = \omega_x/\omega_z$ and $\beta = \omega_y/\omega_z$. We take $\alpha$ = $\beta$ = 0.5 with $\omega_o \equiv \omega_z = 25\times 2\pi \mathrm{Hz}$ and form the torus by including a core potential with parameters $V_c=144$, and $\sigma_c=1.88$ with $\rho^2 = x^2 + y^2$. The ground state chemical potential is $\mu=7.6$ for $N=2.5x10^5$ $^{23}\mathrm{Na}$ atoms, corresponding to a nonlinear scaling value $g\ N=2028$.

During the dynamical evolution, we use a time-dependent barrier potential of the form:
\begin{equation}
V_\mathrm{b}(\rv,t)=f(t)  {{V_\mathrm{s} }} V_{\mathrm{bx}}(x)V_{\mathrm{by}}(y) V_{\mathrm{bz}}(z)/8,
\end{equation}
with $f(t)=t/t_\mathrm{r}$ ($f(t)=1$ for
$t>t_\mathrm{r}$) and $V_{\mathrm{bx}}=\tanh(\frac{x-R_x+x_0}{b_\mathrm{s}})+\tanh(\frac{-x+R_x+x_0}{b_\mathrm{s}})$.
Here $R_x$ is the $x$-shift of the center of the barrier while its width is $w_x\sim 2 x_0$. $V_{\mathrm{by}}(y)$ and $V_{\mathrm{bz}}(z)$ have the same form as $V_{\mathrm{bx}}$. The final height of the barrier is $V_\mathrm{s}$ as long as $x_0,y_0,z_0\gg b_\mathrm{s}$.

Before starting the dynamics, we put the condensate in motion by imprinting an appropriate spatially dependent phase $\theta(\rv)$ on the wavefunction. In the waveguide case, $\theta(\rv) =m v y/\hbar$ generates a uniform flow of velocity $v$ along the $y$ direction. In the torus, $\theta(\rv) =m l \phi /\hbar $, with $l$ integer, generates a tangential flow of speed $v(r)=l/\rho$, where $\phi=\arctan(y/x)$.

The results presented in this manuscript are obtained by solving numerically the GPE. After finding the ground state with an imaginary-time propagation, we perform the dynamical evolution in real time. For all the simulations performed we used a finite-element discrete variable representation (DVR)\cite{Collins_2006}. In the waveguide geometry, we employed a split-operator method, with a Fast-Fourier-Transform algorithm used for the kinetic part of the evolution, while for the simulations in the torus we used a Real Space Product Formula \cite{Collins_2006}. The spatial grid for the waveguide calculations was $n_x=60,n_y=120,n_z=60$ (or $n_x=180,n_y=120,n_z=180$ for Figs.~\ref{wg2} and \ref{wg2bis}) with box sizes $L_x=12,L_y=24,L_z=12$ for the axially symmetric case (see below), $n_x=90,n_y=120,n_z=80$ with box sizes $L_x=18,L_y=24,L_z=16$ for the axially asymmetric case, while the time-step was $dt=10^{-4}$. In the torus simulations, we used a spatial grid consisting of $80$ elements in each dimension with $4$ DVR Gauss-Legendre functions in each element spanning
cubic box lengths of $[-12,12]$, $[-12,12]$, and $[-10,10]$ in the $x$, $y$, and $z$-directions respectively. This choice gave $241$ grid points in each direction. We also tested the convergence with $5$ basis functions and $321$ points with only a few percent change in basic quantities such as energies, momentum, and positions.

{\bf Criterion for instability}
\label{sec:criterion}

\begin{figure}
  \includegraphics[scale=.8]{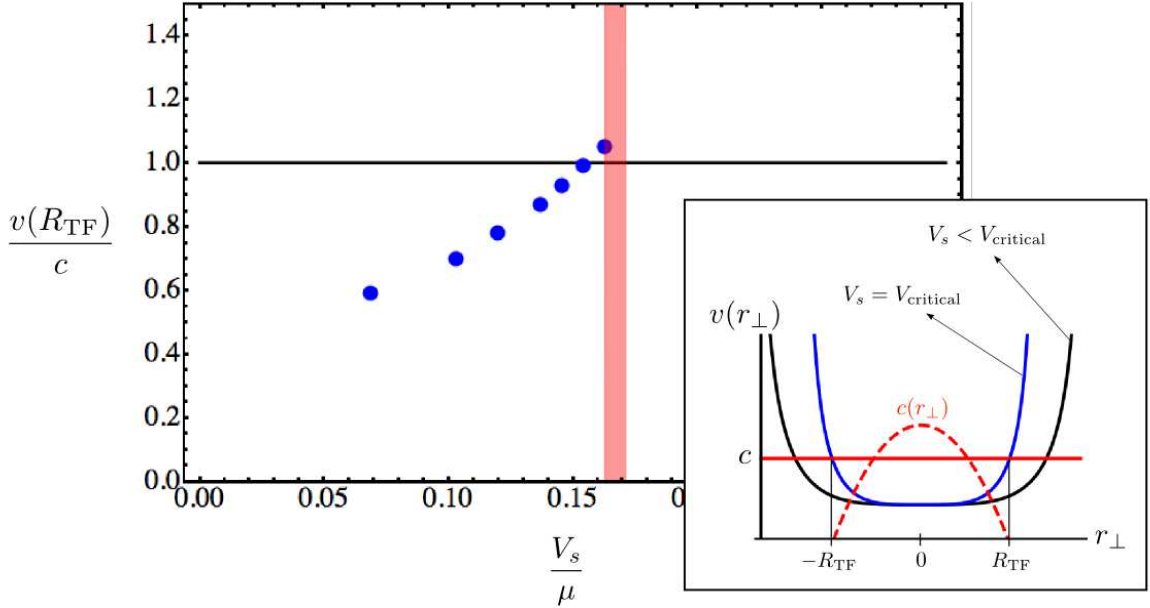}
  \caption {
        (color online). Ratio of the higher local fluid velocity at the Thomas-Fermi radius, $v(R_{TF})$, to the sound speed $c$ inside the barrier, as a function of the barrier height. The red-shaded area correspond to the critical point plus uncertainty. The results are obtained for the cylindrically symmetric waveguide $\gamma=1$ and initial flow velocity $v=1.05$. The inset shows a sketch of the behavior of the local fluid velocity along a radial cut inside the barrier region. The red solid line indicates the value of the sound speed $c$, which with transverse harmonic confinement is just the average of the local sound speed, $c(r_\perp)$ (red dashed line) over the transverse plane. The black(blue) solid line corresponds to a subcritical(critical) condition.}
\label{criterion}
\end{figure}

In the hydrodynamic regime, when the healing length $\xi$ is much smaller than all other length scales (the smallest of which is the barrier width along the flow direction), we find that the onset of instability coincides with a simple condition: inside the barrier region, the local fluid velocity at the classical (Thomas-Fermi) surface of the cloud equals the sound speed, as depicted in the inset of Fig.~\ref{criterion}. 
It is important to precisely define the sound speed which sets the threshold for the critical velocity. The latter is the Bogoliubov sound speed, calculated inside the barrier region, for the low-lying modes propagating along the flow, taken as if the system was homogeneous in this direction. In the case of a waveguide with a harmonic transverse confinement, and within the Thomas-Fermi approximation, this sound speed is simply the average of the local sound speed on a plane perpendicular to the flow $c=c(0)/\sqrt{2}$, where $c(0)$ is the local sound speed at the center of the transverse harmonic trap \cite{Zaremba_1997}.

We verified this numerically in the waveguide case, as shown in Fig.~\ref{criterion}. In a waveguide with cylindrical symmetry, this happens simultaneously at points on a circle perpendicular to the flow, from where a vortex ring will then enter. In the torus, due to a higher flow speed at the inner edge of the annulus, the critical condition is reached first on the interior of the cloud, with the consequence that vortex rings, if ever formed, must be asymmetric, as we discuss below.

Slightly below the critical point, we observe vortices getting closer to the  edges of the cloud but failing to enter. The presence of these ``ghost'' vortices \cite{ghost,Piazza_2009} suggests that a pre-instability is triggered at the edges of the condensate, but, since the vortices do not enter the bulk region, it is not sufficient to dissipate the superflow.

In a previous work \cite{Piazza_2009}, we verified the same condition for instability to hold for a two-dimensional toroidal BEC. The results reported here represent an extension to a three-dimensional case. As discussed above, this criterion allows for a simple understanding of the details of vortex penetration dynamics, based on the observation that vortex cores enter first where the critical condition if first reached.

{\bf Instability dynamics in the waveguide}

In analogy with the standard situation where a superfluid flows through a channel, we consider a waveguide with periodic boundary conditions along the flow direction $y$, and the barrier creating the equivalent of the channel. This situation can be realized experimentally with a BEC inside an elongated trap along the flow direction. As stated above, before raising the barrier in the simulations, a stationary flow is created by imprinting a phase $ m v y/\hbar$ on the condensate wavefunction, where $v$ is the constant flow speed. This corresponds in the experiment to sweeping the barrier across the cloud at a constant velocity. We consider a symmetric harmonic confinement in the transverse $x-z$ plane, such that $d_x=\sqrt{\hbar/m\omega_{x}}$ and $d_z=\sqrt{\hbar/m\omega_{z}}$ are both sufficiently larger than the bulk healing length, so that the effect of transverse degrees of freedom comes into play, giving rise to a fully three-dimensional dynamics. Such a setup has been used in \cite{Steinhauer_2009}, where, moreover, the possibility of creating a penetrable repulsive barrier with a control over a length comparable to the healing length has been demonstrated.

\begin{figure}
  \includegraphics[scale=0.53,clip]{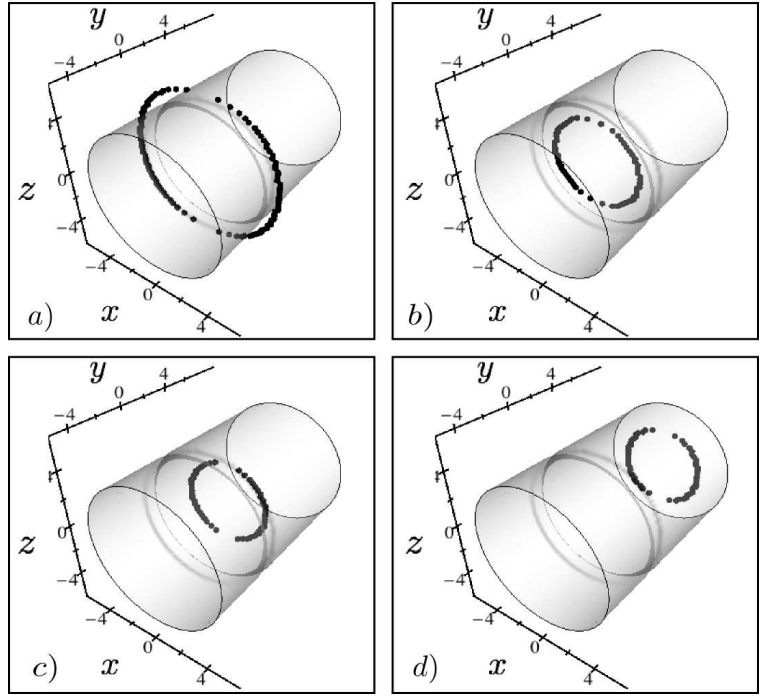}
  \caption {
        (color online). Four subsequent stages of the vortex ring penetration in the waveguide. The gray surface indicates the position of the Thomas-Fermi surface of the condensate. Black dots show the position of the vortex cores. Here the waveguide is axially symmetric $\gamma=1$, the initial flow velocity is $v=1.05$, and the final barrier height is $V_s=0.17\mu$.  a) The ring is shrinking around the cloud in barrier region, still outside the Thomas-Fermi borders of the cloud. b) The ring has just entered the cloud. c) The ring has shrunk to its final size and is already outside the barrier region, moving along the flow direction. d) The ring has moved far from the constriction region, with a constant speed and radius.}
\label{wg1}
\end{figure}

In this configuration, as soon as the critical barrier height is reached, a vortex ring detaches from the system boundaries and starts shrinking into the cloud inside the barrier region, as shown in Fig.~\ref{wg1}, a). 
In the figure, black points indicate the position of vortex cores while the gray surface corresponds to the Thomas-Fermi surface of the cloud.
Detection of vortex cores is performed using a plaquette method, described in \cite{Foster_2010}.

Depending on the initial flow velocity, waveguide transverse section, and barrier height at the critical point, the ring attains a certain radius and velocity with which it propagates in the flow direction in a stable fashion, as long as axial symmetry is preserved, as depicted in Fig.~\ref{wg1}, c) and d). Here the vortex ring eventually propagates at the speed $u_{\rm r}=0.65$, and a radius $R=2.9$. 

For sufficiently strong barriers, the ring shrinks to a point crossing the transverse section, thereby annihilating and completing a full single phase-slip, as shown in Fig.~\ref{wg2}. After this process, the velocity has dropped everywhere by the same quantized amount. We observe that such an event is the three-dimensional analogue of vortex-anti-vortex annihilation in two dimensions \cite{Piazza_2009}. 
\begin{figure}
  \includegraphics[scale=0.6,clip]{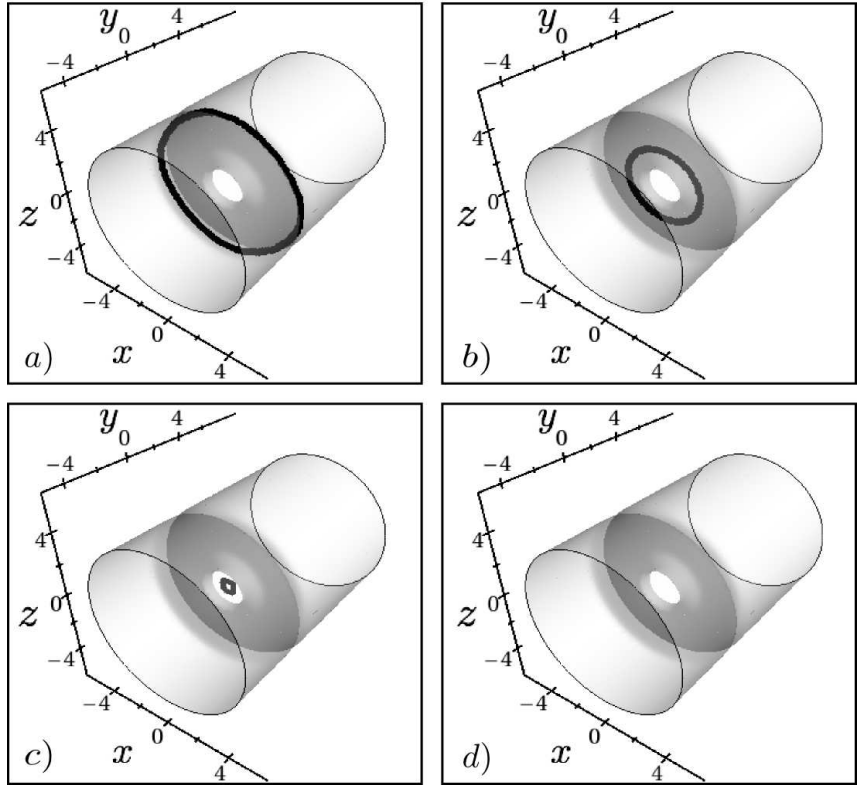}
  \caption {
    (color online). Four subsequent stages of the vortex ring annihilation in the waveguide. The gray surface indicates the position of the Thomas-Fermi surface of the condensate. Black dots show the position of the vortex cores. Here the waveguide is axially symmetric $\gamma=1$, the initial flow velocity is $v=0.52$, and the final barrier height is $V_s=0.94\mu$. a) The ring is shrinking around the cloud in the barrier region, still outside the classical borders of the cloud. b) The ring has entered the cloud. c) The ring is about to shrink completely and annihilate. d) The ring has annihilated and no vortex core is now inside the Thomas-Fermi surface.
     }
\label{wg2}
\end{figure}
It is interesting to analyze the ring annihilation event in more detail. In Fig.~\ref{wg2bis} a) and b), respectively, we show the position of the vortex cores toghether with the density on a plane parallel to the flow direction, at a time just before and just after the ring has shrunk to a point. In Fig.~\ref{wg2bis} c), the density on the same plane is plotted at four subsequent times after the annihilation.
The details of self-annihilation process which we observe are consistent with the previous studies of axisymmetric vortex ring solutions \cite{roberts_1982, roberts_1986, roberts_2004, pitaevskii_2004, stringari_book}. Namely, if we consider the position of the vortex cores, we see that the points of phase singularity form a loop which shrinks inside the constriction, Fig.~\ref{wg2bis} a), and whose radius eventually becomes zero, Fig.~\ref{wg2bis} b). At this moment, we see the zero-density core being filled with atoms, thereby transforming into a density depression, which further propagates out of the constriction as a rarefaction pulse, as shown in Fig.~\ref{wg2bis} c). Finally, this rarefaction pulse decays into sound.
\begin{figure}
  \includegraphics[scale=0.6,clip]{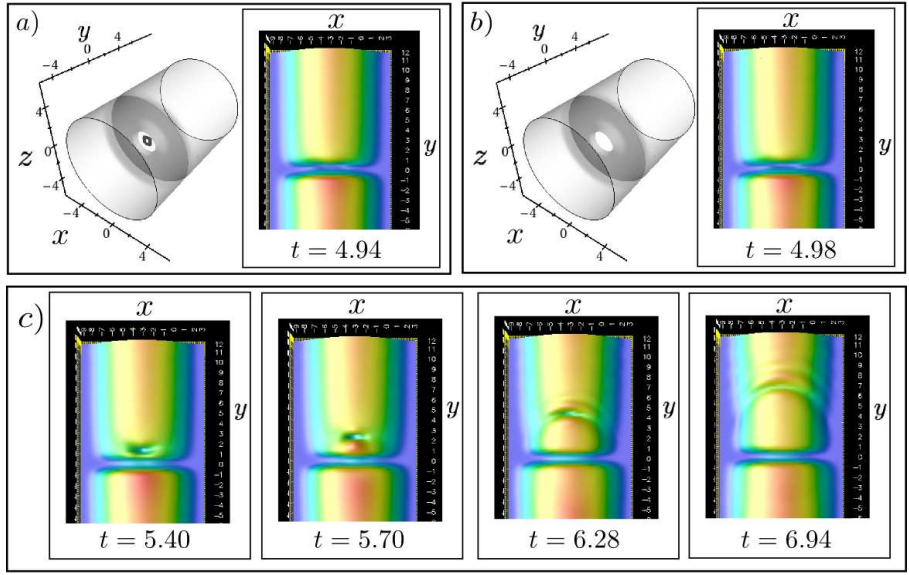}
  \caption {
    (color online). Details of the ring self-annihilation event studied in Fig.~\ref{wg2}, at six subsequent times. In a) and b), left panel, the gray surface indicates the position of the Thomas-Fermi surface of the condensate, while black dots show the position of the vortex cores. In a) and b), right panel, and c), the density on a $z=0$ plane parallel to the flow direction is shown. A very small loop structure of vortex cores, in a), shrinks to a point and has disappeared in b). The ring has transformed into a rarefaction pulse, whose propagation and decay into sound appears in c).
     }
\label{wg2bis}
\end{figure}

Both in Figs.~\ref{wg1} and \ref{wg2}, we see that, at least initially, the vortex cores are subjected to an essentially radial motion, giving rise to the shrinking of the vortex ring.
This behavior can be understood by considering what contributes to the motion of vortex cores in non-homogeneous systems \cite{Pethick_2006}. Let us consider a single vortex core located at some position $x_c$. Its velocity would be the sum of two terms: i) the background flow velocity at $x_c$, and ii) a term perpendicular to the gradient of the density at $x_c$. Both contributions must be calculated as if the vortex was not present. Since the relative weight of term ii) is proportional to the value of the healing length calculated at $x_c$, in high density regions, a vortex core will move mainly with the background superfluid velocity, while in low density regions, it will move mainly due to the gradient of the density. 
Therefore, when the vortex ring is inside the constriction, where the density is low, and especially when it is close to the Thomas-Fermi surface, it will mainly move perpendicular to the gradient of the density, whose main contribution comes from the density modulation induced by the barrier along the flow direction. This results in an essentially radial motion of the cores, and thus into the shrinking of the ring.

{\bf Instability dynamics in the torus}

\begin{figure}
  \includegraphics[scale=0.8,clip]{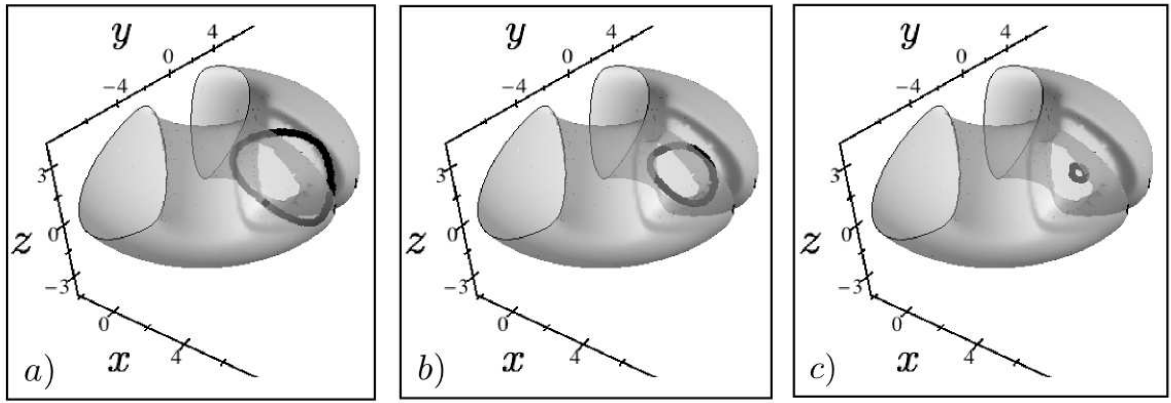}
  \caption {
        (color online). Three subsequent stages of the vortex ring annihilation in the torus. The gray surface indicates the position of the Thomas-Fermi surface of the condensate. Black dots show the position of the vortex cores. Here the initial circulation is $l=1$, and the final barrier height is $V_s=0.5\mu$. a) The ring is shrinking around the cloud in barrier region, still outside the classical borders of the cloud. b) The ring has entered the cloud. c) The ring is about to shrink completely and annihilate.}
\label{tor3}
\end{figure}

In the toroidal geometry, the scenario is richer since the tangential flow velocity at a given total angular momentum, due to the conservation of quantized circulation, decreases like 1/r, where r is the distance from the center of the torus. This introduces an asymmetry between the inner and the outer edges of the cloud, which is not present in the waveguide case. Consequences of this asymmetry on vortex nucleation in two dimensions are discussed in \cite{Piazza_2009}. 
In three-dimensional configurations, the result is that vortex rings are transient features in a toroidal geometry.

\begin{figure}
  \includegraphics[scale=0.6,clip]{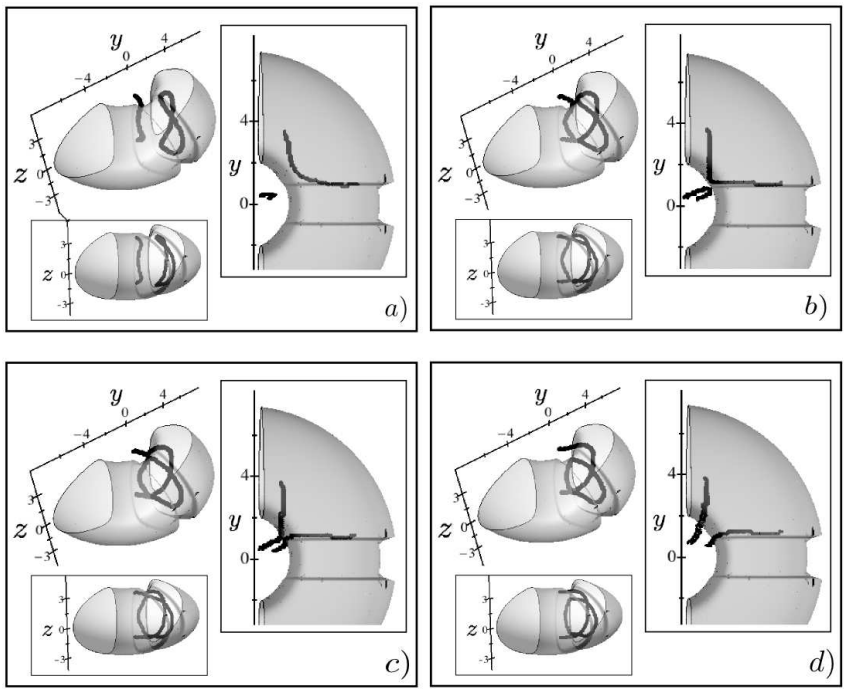}
  \caption {
        (color online). Four subsequent stages of the vortex ring breaking in the torus. The gray surface indicates the position of the Thomas-Fermi surface of the condensate. Black dots show the position of the vortex cores. Insets show the top and side views. Here the initial circulation is $l=4$, and the final barrier height is $V_s=0.2\mu$. a) The vortex ring is bending to form a right angle. b) The vortex ring has formed a right angle whose vertex is close to a vortex line coming frome the center of the torus. c) The vortex ring and line have just reconnected: a vortex line and a portion of a ring vortex are now inside the Thomas-Fermi surface. d) The vortex line and the ring have moved apart. 
      }
\label{tor2}
\end{figure}
 
The ring vortex either breaks or shrinks to a point and annihilates for sufficiently strong barriers. 
While the annihilation process, shown in Fig.~\ref{tor3}, is very similar to the one taking place in a waveguide,
the vortex ring breaking mechanism reflects instead more evidently the asymmetry in the velocity field. 

As shown in Fig.~\ref{tor2}, the ring is strongly deformed since the inner part moves faster (the velocity of a vortex core, when the healing length is much smaller than the length scale of density variation, is essentially given by the background flow velocity \cite{Pethick_2006}). The deformation increases in time up to the point at which the ring bends in on itself, forming a right angle at whose vertex a kink is present, as shown in Fig.~\ref{tor2} b). Meanwhile, a vortex line coming from the inner core of the torus approaches the vertex, also forming a kink in correpondence to the latter. Eventually, the line and the ring connect, joining each other at the position of the kinks. Such an event produces a vortex line plus a new ring vortex, appearing in Fig.~\ref{tor2} c).
Thus, the breaking is actually a vortex reconnection. Vortex reconnections play an important role in turbulent scenarios, being a very efficient mechanism for increasing the number of vortices in the system \cite{Vinen_2002}. 
When the barrier is not strong enough to make both the edges of the annulus unstable, we also observed cases in which the ring is not formed at all, and a strongly bent vortex line enters the cloud from its inner edge, to circulate around the torus. 
The fact that the vortex line enters the inner edge of the annulus is due to the above mentioned asymmetry in the velocity field, decaying as the inverse of the distance from the center of the torus, which makes the instability set in there first, according to the criterion discussed in section \ref{sec:criterion}.

\begin{figure}
  \includegraphics[scale=0.8,clip]{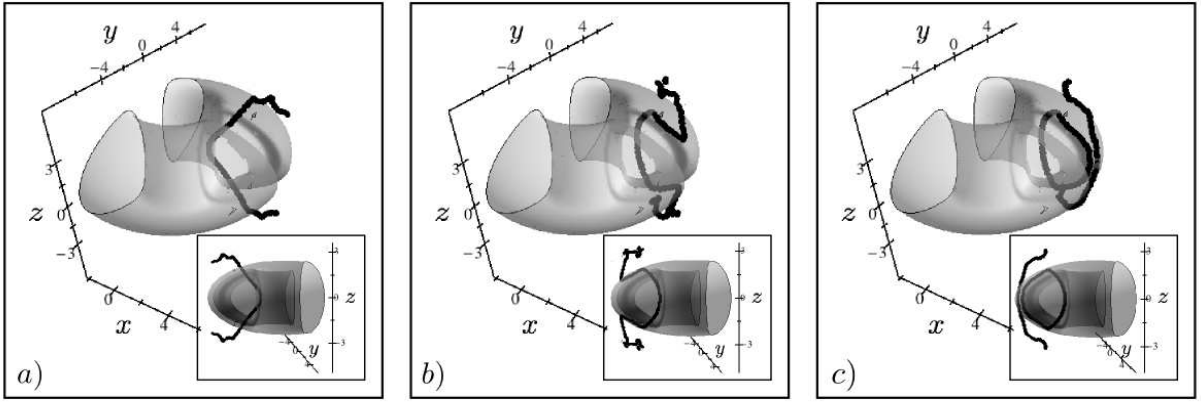}
  \caption {
        (color online). Three subsequent stages of the vortex ring formation in the torus. The gray surface indicates the position of the Thomas-Fermi surface of the condensate. Black dots show the position of the vortex cores. Insets show the side view. Here the initial circulation is $l=4$, and the final barrier height is $V_s=0.2\mu$. a) The vortex line is bending around the cloud in barrier region, partially outside the Thomas-Fermi borders of the cloud. b) The vortex line has developed two kinks. c) The vortex ring has just formed.}
\label{tor1}
\end{figure}

The formation of a vortex ring in the torus is very interesting, since it can be seen as a reconnection of a vortex line with itself. 
As shown in Fig.~\ref{tor1}, a bended vortex line is always present at first. While the bending increases, the vortex line develops two sharp kinks whose tips get closer to each other, up to when they join, thereby cutting the original line into a vortex ring plus a yet another line. The latter is then reabsorbed at the system's boundary. 

{\bf Connection between effective two- and three-dimensional geometries}

\begin{figure}
  \includegraphics[scale=0.64,clip]{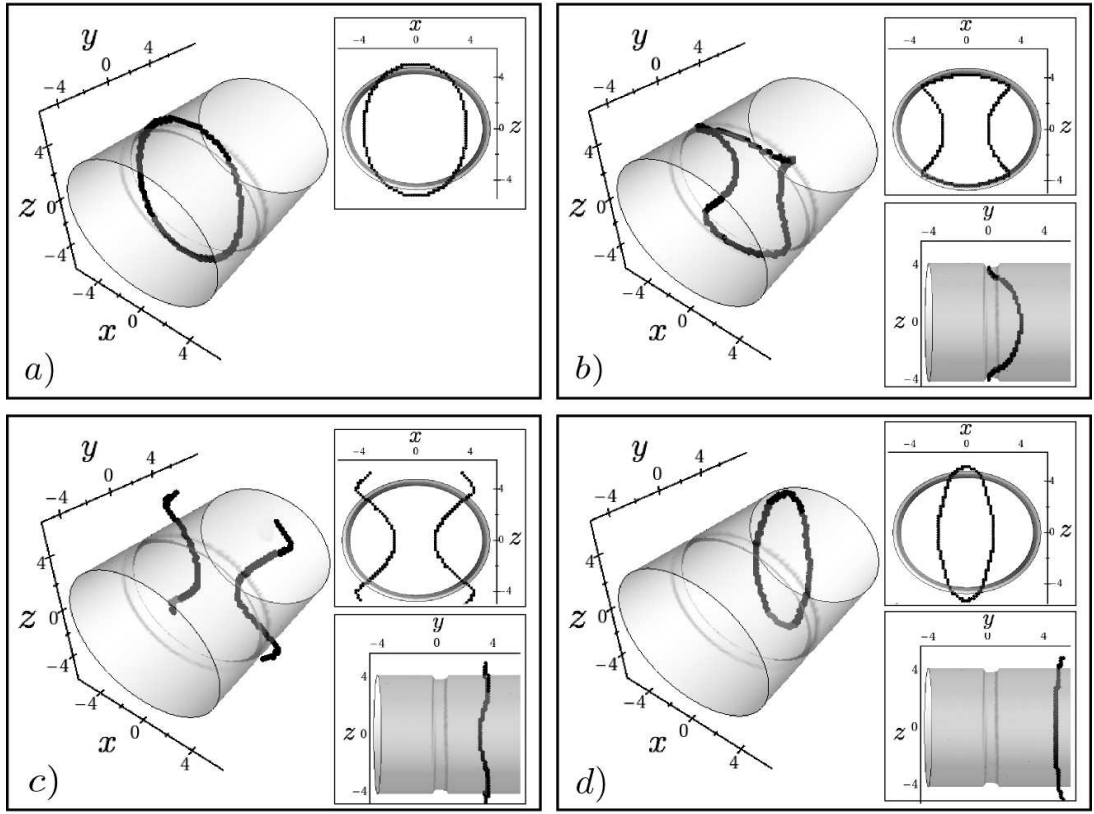}
  \caption {
        (color online). Four subsequent stages of the vortex ring dynamics in the axially asymmetric waveguide. The gray surface indicates the position of the Thomas-Fermi surface of the condensate. Black dots show the position of the vortex cores. Insets show the front and side view. Here the waveguide is non-axially symmetric $\gamma=1.2$, the initial flow velocity is $v=1.05$, and the final barrier height is $V_s=0.13\mu$. a) A partially-ghost ring vortex has formed. b) The ring vortex is strongly deformed. c) The ring vortex has broken up leaving two vortex lines. d) The vortex lines have joined back to form a new ring vortex. }
\label{wgsq}
\end{figure}

When the condensate is effectively in two dimensions, meaning that the cloud size along the third direction is comparable with the healing lentgth $\xi$, vortex lines oriented along the direction of tighter confinement can become the preferred excitations with respect to vortex rings \cite{Collins_2000}. In studying a condensate inside an effective two-dimensional toridal trap (the system's size along the axis of the torus was about one healing length) \cite{Piazza_2009}, we indeed observed that superfluid flow dissipation took place through the formation of vortex lines entering the cloud.
For the trap configurations considered in this manuscript, we have observed vortex ring formation in the very low density regions outside the condensate classical Thomas-Fermi surface. As suggested above, we could call these ghost \cite{ghost} vortices since they are not visible from the condensate density profile. In some cases, as discussed above, these ghost vortex rings are able to enter completely the classical surface of the cloud, transforming into what we should then call ``real'' ring vortices.

In the crossover between effective two- and three-dimensional regimes, moving from a scenario in which only vortex lines are present to one in which real vortex rings come into play, it is reasonable to expect an intermediate regime in which ghost vortex rings are formed, but the condensate is sufficiently squashed along the third dimension that the full vortex loop is not able to enter the cloud's classical surface. In this regime, superfluid would be dissipated by vortex rings which are partly real and partly ghost, appearing as simple bent vortex lines in the density profile.

An example of such situation is given in Fig.~\ref{wgsq}, where the waveguide is non-axially symmetric about the flow direction ($\gamma=1.2$). After the instability sets in, a ghost vortex ring forms and shrinks around the cloud in the barrier region up to when we observe a full-fledged ring vortex, which is only partially inside the Thomas-Fermi surface of the condensate (Fig.~\ref{wgsq}a)). However, since the part of the ring which is inside the Thomas-Fermi surface moves with a larger speed along the flow direction with respect to the part which remains outside, the ring vortex is soon deformed (Fig.~\ref{wgsq}b)), and eventually breaks up (Fig.~\ref{wgsq}c)). The vortex cores located inside the Thomas-Fermi surface of the cloud move, to a good approximation, along with the background velocity field. On the other hand, cores in the low density region outside the surface of the cloud essentially do not feel the background velocity and move along the flow direction only because of the presence of transverse density gradients. After the vortex ring breaks up, the two lines move downstream and continue to deform, and eventually re-join to form a vortex ring (Fig.~\ref{wgsq}d)). The latter undergoes the same deformation described above, leading to another break up.

We also verified that, even in a very slightly non-axially symmetric waveguide, ring vortices eventually break up. With $\gamma=1.05$, the deformation is created more slowly with respect to the $\gamma=1.2$ case, but the ring breaks up anyway, though at a later time.

{\bf Conclusions}

We studied the superfluid instability condition and dynamics in the presence of a three-dimensional constriction across the flow. We solved the GPE, which would accurately describe an ultracold dilute Bose gas, but would anyway provide a useful insight for the understanding of the instability in strongly correlated superfluids like Helium II.
We considered two different setups: a waveguide, which could be experimentally realized by trapping an ultracold Bose gas inside a cigar-shaped trap, and a torus. We found that the instability sets in when the fluid velocity at the Thomas-Fermi surface of the cloud equals the sound speed inside the constriction region. This critical condition explains the strong dependence of the instability dynamics on the geometry considered, and in particular on the asymmetries present in the flow velocity and density. 
In an axially-symmetric waveguide, the superfluid dissipates through the nucleation of circular ring vortices that shrink across the flow within the barrier regions. As soon as the axially symmetry is broken, the rings are strongly deformed and can break up into vortex lines or not form at all.
The phase slip instability observed here can be considered as a superfluid flow dissipation mechanism, since it subtracts energy from the potential flow of the superfluid. However, GPE does not include the dissipative processes which bring the system to thermal equilibrium. In a realistic situation, the vortices, carrying the energy subtracted from the superfluid, are supposed to eventually thermalize, with the result of heating the system. The effect of dissipation on vortex nucleation has been studied by adding a dissipation term to the GPE \cite{Biao_2010}. The role of finite temperature on the superfluid instability scenario presented here deserves further study. 

{\bf Acknowledgements}. 

We are acknowledge many helpful discussions with L. P. Pitaevskii.
We especially thank Chris Ticknor for bringing to our attention the plaquette technique, and providing
us with subroutines for its implementation within our analysis programs. 
The Los Alamos National Laboratory is operated by Los Alamos National Security, LLC for the National Nuclear Security Administration of the U.S. Department of Energy under Contract No.~DE-AC52-06NA25396.

\end{document}